# A Computational Model of Crowds for Collective Intelligence

JOHN PRPIĆ, Beedie School of Business - Simon Fraser University
PIPER JACKSON, Modelling of Complex Social Systems Program - Simon Fraser University
THAI NGUYEN, Computing Science - Simon Fraser University

1. INTRODUCTION

In this work, we present a high-level computational model of IT-mediated crowds for collective intelligence. We introduce the Crowd Capital perspective as an organizational-level model of collective intelligence generation from IT-mediated crowds, and specify a computational system including agents, forms of IT, and organizational knowledge.

2. IT-MEDIATED CROWDS FOR COLLECTIVE INTELLIGENCE

A new and rapidly emerging paradigm of socio-technical systems for collective intelligence can be found in the increasing use of IT-mediated crowds by organizations for knowledge purposes. In this domain, Crowdsourcing [Brabham 2008, 2010] is being widely applied in a growing number of contexts, and the knowledge generated from these phenomena is well-documented [Huberman 2008, Huberman et al. 2009]. At the same time, other organizations are generating collective intelligence by putting in place IT-applications known as Predication Markets [Hankins and Lee 2011, Arrow et al. 2008], which serve to gather large sample-size forecasts from distributed populations. Similarly, we are also observing many organizations using IT-tools such as "Wikis" [Majchrzak 2006, Majchrzak et al. 2013] to access the knowledge of dispersed populations within the boundaries of their organization. Further still, other organizations are implementing Citizen Science techniques [Crowston and Prestopnik 2013] to accumulate scientific research knowledge from the public at large through IT. In a related manner, many organizations are using IT-mediated crowds as partners for innovation [Boudreau and Lakhani 2013].

Among all of these examples of different socio-technical systems currently being implemented by organizations, we see that organizations are engaging crowds through IT, for the purpose of generating knowledge [Prpić and Shukla 2013, 2014]. Generating knowledge in this way is a prime example of "groups of individuals doing things collectively that seem intelligent" [Malone et al. 2009].

3. THE CROWD CAPITAL PERSPECTIVE

The Crowd Capital perspective [Prpić and Shukla 2013, 2014] is an organizational-level model outlining how and why organizations are using IT to engage crowds for knowledge purposes. The Crowd Capital perspective captures the essence and dynamics of the numerous substantive research areas already mentioned here including: prediction markets, wikis, citizen science, crowdsourcing and open innovation platforms, and formulates a generalized model of knowledge generation from IT-mediated crowds (see Figure #1 below).

**Figure #1** – The Crowd Capital Perspective[1]

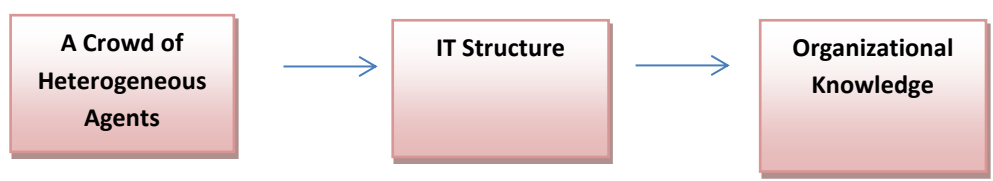

---

[1] Adapted from Prpić and Shukla [2013; 2014}

## 4. THE COMPUTATIONAL MODEL

Computational modelling is a scientific method of describing phenomena using models that are capable of being run as programs on a computer. This implies that such a model is mathematical in nature, since computers are only capable of performing operations which are fundamentally mathematical. It also necessitates explicitness with regard to the assumptions underlying the model, since vagueness will prevent the model from being executed [Börger and Stärk 2003]. Computational modelling allows us to examine our understanding of phenomena in a rigorous and interactive manner, and is particularly appropriate for considering complex topics such as social systems [Epstein 1999].

A complete specification of the Crowd Capital computational model will be presented in forthcoming works, including a variety of useful simulation experiments illustrating the dynamics of the system. In the remainder of this work, we introduce the general structure of the Crowd Capital model and its elements.

The three elements presented in Figure #1 interact in the following manner:

1. If an individual member of the crowd (an agent) is currently motivated to participate, they request a task from the IT structure. In the case of a collaborative IT structure (see below), the task may already be partially completed.

2. After work is completed on a task, the task is sent back to the IT structure, and contains an overall completion-level between 0.0 (nothing) and 1.0 (perfect).

3. The IT periodically updates the organizational knowledge base. This can be as simple as passing along the completion levels of tasks, or can involve a more complicated algorithm for filtering submissions.

The Crowd Capital computational model views a crowd as being a particular collection of independently-deciding individuals [Reiter and Rubin 1998, Surowiecki 2005], here modelled as heterogeneous human agents who have:

   a) Motivation

   b) Differing task-completion success rates

These characteristics allow us to model individual crowd members in respect to their manner of interaction with an IT-mediated knowledge generation system. Motivation determines the activity level of an agent in regard to the IT structure (i.e. do they interact hourly, daily, or yearly with the IT structure?). Success rates determine a crowd member's productivity in regard to a knowledge generation activity. In this way, we are able to model individuals along the spectrum from novice to expert, which is vital when modelling a heterogeneous participant-base that comprises a crowd.

The Crowd Capital computational model views IT structure as occurring in either an episodic or collaborative form, premised on the interface that is coded into the artifact. Collaborative forms of IT structure allow and necessitate interaction among the agents through the IT for a knowledge resource to form (e.g., a wiki or social media), while episodic forms of IT structure do not allow nor necessitate interaction among agents for a knowledge resource to form (e.g., reCAPTCHA or Foldit). In Crowd Capital computational model, knowledge tasks are embedded in the particular form of IT structure and include:

   a) A level of difficulty

   b) A unique signifier

   c) A cost to process by the organization

d) A level of completion

The Crowd Capital computational model views organizational knowledge as the system outcome with the following metrics:

a) Number of submissions

b) Number of submissions completed

These metrics are vital in determining the effectiveness of the knowledge generation system for a given scenario. Other metrics are of course possible, depending on the nature of the task being considered, and an organization's prerogatives. However, these two are sufficient for a high-level model showing the general process of organizational knowledge generation from crowds through IT. The first two sets of system characteristics, i.e., those related to users and tasks, are scenario parameters. As such, they are set for a given scenario at the outset, and determine how a given experiment will unfold. The organizational knowledge metrics are used to measure the system behaviour, and can be thought of as the output values of each experiment.

5. CONCLUSION

In this work we report on research in progress that has developed a computational model of IT-mediated crowds for collective intelligence. We introduce the Crowd Capital perspective as an organizational-level model of collective intelligence generation from IT-mediated crowds, and specify a computational system including agents, forms of IT structure, and organizational knowledge.

As far as we are aware, this work, and our simulation software, represents the first attempt to computationally model the organizational knowledge generating capacity of IT-mediated crowds. Further, as far as we know, the work is also the first of its kind to computationally model episodic and collaborative IT structures, distinguishing those forms of IT that require crowd collaboration for knowledge generation, from those that do not.

Our future research will proceed in a number of directions building on the platform of this work and our simulation software. Our goal is a more highly specified computational system, which includes real-world data to populate the elements outlined here.